\documentstyle[preprint,aps]{revtex}

\begin{document}


\draft

\title{
Hydrodynamic modes and pulse propagation in a cylindrical Bose
gas above the Bose-Einstein transition}

\author{T. Nikuni}
\address{Department of Physics, Tokyo Institute of Technology,
Meguro, Tokyo 152, Japan}
\author{A. Griffin}
\address{Department of Physics,
University of Toronto, Toronto, Ontario M5S 1A7, Canada}

\date{\today}
 
\maketitle

\begin{abstract}
We study hydrodynamic oscillations of a cylindrical Bose gas above the
Bose-Einstein transition temperature using the hydrodynamic equations
derived by Griffin, Wu and Stringari.
This extends recent studies of a cylindrical Bose-condensed gas at
$T=0$.
Explicit normal mode solutions are obtained for non-propagating solutions.
In the classical limit, the sound velocity is shown to be the same
as a uniform classical gas.
We use a variational formulation of the hydrodynamic equations to discuss
the propagating modes in the degenerate Bose-gas limit and show there is
little difference from the classical results.
\end{abstract}

\bigskip


\clearpage

\narrowtext
\section{introduction}
Recently Andrews {\it et al}. \cite{andrews}  reported a measurement of a
sound pulse propagation of Bose-condensed cloud in a highly asymmetric
cigar-shaped trap.
They have measured the sound velocity in a
Bose-condensate as a function of the density.
Zaremba \cite{zaremba} gave a detailed analysis of the collective excitations
in a cylindrical Bose gas starting from the $T=0$ quantum hydrodynamic
equation of Stringari~\cite{st1}, which is based on the Thomas-Fermi
approximation.
The sound velocity of a condensate pulse
obtained in Ref.~\cite{zaremba} is in good agreement
with the experimental observations \cite{andrews}.
More recently, several other theoretical studies have discussed the
excitations and pulse propagation in a cigar-shaped
trap at $T=0$~\cite{st2,pethick,machida}.

In this paper, we consider the analogous modes in a hydrodynamic regime 
(where collisions ensure local thermal equilibrium) for
a cylindrical Bose gas above the Bose-Einstein transition temperature
$T_{\rm BEC}$.
Our starting point is the hydrodynamic equations derived by Griffin,
Wu and Stringari \cite{gws}.

\section{Hydrodynamic normal mode equations for a
cylindrical Bose gas}
The linearized hydrodynamic equation for the velocity
fluctuations ${\bf v}({\bf r},t)$ derived by Griffin, Wu and
Stringari is~\cite{gws,kavou}: 
\begin{equation}
m\frac{\partial^2{\bf v}}{\partial t^2}=
\frac{5P_0({\bf r})}{3n_0({\bf r})}\bbox{\nabla}(\bbox{\nabla}\cdot
{\bf v})
-\bbox{\nabla}[{\bf v}\cdot \bbox{\nabla}U_0({\bf r})]
-\frac{2}{3}(\nabla\cdot{\bf v})\bbox{\nabla}U_0({\bf r})
-\frac{\partial}{\partial t}\bbox{\nabla}\delta U({\bf r},t),
\label{eq1}
\end{equation}
where $U_0({\bf r})$ is the static cylindrical trap potential and
$\delta U({\bf r},t)$ is a small time-dependent external perturbation.
The equilibrium local density $n_0({\bf r})$ and the equilibrium
local kinetic pressure $P_0({\bf r})$ in (\ref{eq1}) are given by
\begin{equation}
n_0({\bf r})=\frac{1}{\Lambda^3}g_{3/2}(z_0),\ 
P_0({\bf r})=\frac{k_{\rm B}T}{\Lambda^3}g_{5/2}(z_0),
\label{eq2}
\end{equation}
where $z_0({\bf r})\equiv e^{\frac{1}{k_{\rm B}T}(\mu_0-U_0)}$
is the local equilibrium fugacity, $\Lambda\equiv(2\pi\hbar^2/mk_{\rm B}T)
^{1/2}$ is the thermal de Broglie wave-length and $g_n(z)=\sum_{l=1}^{\infty}
z^l/l^n$ are the well-known Bose-Einstein functions.

Throughout this paper, we shall limit our discussion to a purely cylindrical
harmonic trap potential
\begin{equation}
U_0({\bf r})=\frac{1}{2}m\omega_0^2(x^2+y^2).
\label{eq3}
\end{equation}
Inserting this into (\ref{eq1}), we obtain coupled equations for the
radial ${\bf v}_{\perp}$ and axial $v_z$ velocity fluctuations
\begin{mathletters}
\begin{eqnarray}
m\frac{\partial^2 {\bf v}_{\perp}}{\partial t^2}
&=&\frac{5P_0}{3n_0}\bbox{\nabla}_{\perp}
(\bbox{\nabla}_{\perp}\cdot{\bf v}_{\perp})
-m\omega_0^2\bbox{\nabla}_{\perp}({\bf r}_{\perp}\cdot{\bf v}_{\perp})
-\frac{2}{3}m\omega_0^2(\bbox{\nabla}_{\perp}\cdot{\bf v}_{\perp})
{\bf r}_{\perp} \cr
&& \cr
&&+\left(\frac{5P_0}{3n_0}\bbox{\nabla}_{\perp}-\frac{2}{3}m\omega_0^2
{\bf r}_{\perp}\right)\frac{\partial v_z}{\partial z},\\
&& \cr
m\frac{\partial^2 v_z}{\partial t^2}
&=&\frac{5P_0}{3n_0}\frac{\partial^2 v_z}{\partial z^2}
+\left(\frac{5P_0}{3n_0}\bbox{\nabla}_{\perp}
-m\omega_0^2{\bf r}_{\perp}\right) \cdot
\frac{\partial {\bf v}_{\perp}}{\partial z},
\end{eqnarray}
\end{mathletters}

\noindent
where we have set $\delta U({\bf r},t)=0$ since we are interested in normal
mode solutions (driven solutions will be discussed in Section VI).
We use the convention
${\bf v}({\bf r},t)={\bf v}_{\omega}({\bf r})e^{-i\omega t}$.
Finally, we shall look for solutions of the kind
\begin{equation}
 {\bf v}_{\omega}({\bf r})=(xf(r_{\perp}),yf(r_{\perp}),h(r_{\perp}))e^{ikz},
\label{eq5}
\end{equation}
where $r_{\perp}=\sqrt{x^2+y^2}$.
That is, we assume that the functions $f$ and $h$ do not depend on the radial
azimuthal angle.

Using (\ref{eq5}) in (4), one finds, after some algebra,
a coupled set of equations
for the radial function $f$ and the axial function $h$
\begin{mathletters}
\label{eq6}
\begin{eqnarray}
\omega^2 f&=&
-c_0^2(r_{\perp})\left(\frac{\partial^2f}{\partial r_{\perp}^2}
+\frac{3}{r_{\perp}}
\frac{\partial f}{\partial r_{\perp}}\right)
+\omega_0^2\left(\frac{10}{3}f+\frac{5}{3}r_{\perp}
\frac{\partial f}{\partial r_{\perp}}\right)
-ik\left[c_0^2(r_{\perp})\frac{1}{r_{\perp}}\frac{\partial h}
{\partial r_{\perp}}
-\frac{2}{3}\omega_0^2h\right], \\
&& \cr
\omega^2h&=&c_0^2(r_{\perp})k^2h-ik
\left[c_0^2(r_{\perp})\left(2f+r_{\perp}\frac{\partial f}
{\partial r_{\perp}}
\right)-\omega_0^2r_{\perp}^2f\right].
\end{eqnarray}
\end{mathletters}

\noindent
The position-dependent local "sound velocity" $c_0(r_{\perp})$
is defined by
\begin{equation}
c^2_0(r_{\perp})\equiv \frac{5P_0(r_{\perp})}{3mn_0(r_{\perp})}
=\frac{5k_{\rm B}T}{3m}B(z_0),\ \ 
B(z_0)\equiv\frac{g_{5/2}(z_0)}{g_{3/2}(z_0)}.
\label{eq7}
\end{equation}
One can show that the normal mode solutions of (6a) and (6b) satisfy
the following orthogonality 
$$
\int \ d{\bf r}\ n_0({\bf r}){\bf v}^*_{\omega'}({\bf r})\cdot
{\bf v}_{\omega}({\bf r})=0,\ {\rm if}\ \omega_{n'}\neq \omega_n,
\eqno(8{\rm a})
$$
or, more explicitly,
$$
\int_0^{\infty} dr_{\perp}\ r_{\perp}n_0(r_{\perp})
[r_{\perp}^2f^*_{n'}(r_{\perp})f_{n}(r_{\perp})+h^*_{n'}(r_{\perp})
h_{n}(r_{\perp})]=0,\ {\rm if}\ n'\neq n.
\eqno(8{\rm b})
$$
Here the label $n$ specifies the different normal mode solutions.
We remark that while, it is not obvious, if we set the trap frequency
$\omega_0$ to zero, (6a) and (6b) have solutions involving Bessel functions
$J(k_{\perp}r_{\perp})$ with the expected phonon dispersion relation 
$\omega^2=c_0^2(k^2+k_{\perp}^2)$.

\setcounter{equation}{8}

It is convenient to introduce a dimensionless radial variable
\begin{equation}
s\equiv \frac{r_{\perp}^2}{R^2},\ \ R\equiv 
\left(\frac{2k_{\rm B}T}{m\omega_0^2}\right)^{1/2}.
\label{eq9}
\end{equation}
We also introduce a dimensionless frequency and wavevector
\begin{equation}
\bar\omega\equiv\frac{\omega}{\omega_0},\ \ \bar k\equiv kR.
\label{eq10}
\end{equation}
We observe from (2) that $R$ denotes the "size" of the radial density
profile produced by the harmonic potential trap.
In these units, we note $z_0=e^{\beta\mu_0}e^{-s}$ and the classical
density profile $n_0(r_{\perp})\sim e^{-s}$.
In terms of these new dimensionless variables, it is useful to introduce
the new functions:
\begin{equation}
\bar f(s)\equiv-iRf(r_{\perp}),\ \bar h(s)\equiv h(r_{\perp}).
\label{eq11}
\end{equation}
Using (6), the coupled equations for $\bar f$ and $\bar h$ are given by
\begin{mathletters}
\begin{eqnarray}
\bar \omega^2 \bar f&=&\hat L[\bar f]-\bar k\left(\frac{5}{3}B(z_0)
\frac{d\bar h}{ds}-
\frac{2}{3}\bar h\right),\\
\bar \omega^2\bar h&=&
\frac{5}{6}B(z_0)\bar k^2\bar h-\bar k \left[
\frac{5}{3}B(z_0)s\frac{d\bar f}{ds}+
\left(\frac{5}{3}B(z_0)-s\right)\bar f\right],
\end{eqnarray}
\end{mathletters}
\noindent
where we have introduced the operator $\hat L$
\begin{equation}
\hat L[\bar f]\equiv-\frac{10}{3}\left[B(z_0)
s\frac{d^2\bar f}{ds^2}+(2B(z_0)-s)\frac{d\bar f}
{ds}-\bar f\right].
\label{eq13}
\end{equation}
The rest of this paper is based on the equations in (12a) and (12b),
which determine the normal mode velocity fluctuations using (\ref{eq11})
and (\ref{eq5}).
The associated density fluctuations $\delta n({\bf r},t)$ can be found
by using the number conservation law
\begin{equation}
\frac{\partial \delta n}{\partial t}=
-\bbox{\nabla}\cdot(n_0{\bf v}).
\label{eq14}
\end{equation}

\section{Non-propagating modes}

We first examine non-propagating solutions ($k=0$) of (12a) and (12b),
in which case they reduce to the two independent equations
\begin{equation}
\bar \omega^2 \bar f=\hat L[\bar f],\ \ \bar \omega^2\bar h=0.
\label{eq15}
\end{equation}
There is a trivial zero frequency solution 
$\bar f=0,\ \bar h\neq0$, corresponding to
${\bf v}_{\perp}=0,v_z\neq0$.
In this case, the dependence of $\bar h$ on $s$ cannot be determined uniquely.
Using (\ref{eq14}), one finds that $\delta n=0$ for this mode.
The interesting solutions of (\ref{eq15}) with $\bar \omega\neq0$
are given by
\begin{equation}
\bar h=0,\ \ \bar \omega^2\bar f=\hat L[\bar f].
\label{eq16}
\end{equation}
These correspond to oscillations only in the radial direction,
which we shall now discuss.

In the {\it classical} gas limit, the operator $\hat L$ in (\ref{eq13})
simplifies since $B(z_0)=1$.
In this case, we can obtain the complete set of normal mode solutions of
(\ref{eq16}), namely
\begin{equation}
\bar f_n^{(0)}(s)=\frac{1}{n!\sqrt{n}}
\frac{d}{ds}L_n(s),\ \ [\bar \omega_n^{(0)}]^2=\frac{10n}{3},\ \ 
n=1,2,3,\cdots.
\label{eq17}
\end{equation}
Here $L_n(s)$ is the Laguerre polynomial defined by
\begin{equation}
L_n(s)\equiv e^s\frac{d^n}{ds^n}(s^ne^{-s}),
\label{eq18}
\end{equation}
and the orthogonal functions $\bar f_n^{(0)}$ are normalized according to
(see (8b))
\begin{equation}
\int_0^{\infty} ds\ se^{-s}\bar f^{(0)}_n(s)\bar f^{(0)}_{n'}(s)=\delta_{nn'}.
\label{eq19}
\end{equation}
In ordinary variables, the dispersion relation of these $k=0$ solutions
corresponds to 
\begin{equation}
\omega^2_n=n\frac{10}{3}\omega_0^2,\ \ n=1,2,3,\cdots,
\label{eq20}
\end{equation}
and the associated density fluctuation is given by
\begin{equation}
\delta n({\bf r},t)\propto L_n(s=r_{\perp}^2/R^2)\exp(-r_{\perp}^2/R^2)
e^{-i\omega_n t}.
\label{eq21}
\end{equation}
In the lowest $(n=1)$ mode with $\omega_1^2=10\omega_0^2/3$,
$\bar f^{(0)}_1(s)$ is independent of $s$ and $L_1(s)=1-s$.
This is the two-dimensional radial breathing mode.
We note that this particular mode corresponds to one of the
coupled monopole-quadrupole modes found in Ref.~\cite{gws}
for an anisotropic trap in the limit that the axial trap frequency
$(\omega_z)$ is set to zero.
In this same limit $(\omega_z=0)$, the other mode has zero frequency and
a velocity fluctuation is given by ${\bf v}_{\omega}=\alpha(x,y,-5z)$.
This kind of solution is not described by the form (\ref{eq5}) which we are
dealing with in this paper.

For a degenerate Bose gas, in which $B(z_0)$ is now weakly
dependent on $s$ through $z_0=e^{\beta\mu_0-s}$,
one cannot solve (\ref{eq16}) analytically.
We discuss variational solutions in Section V.
However one can check that $\bar f(s)=$constant is a solution, with
frequency $\omega^2=\frac{10}{3}\omega_0^2$.
This is the analogue of the $n=1$ normal mode in (\ref{eq17})
for the classical gas
(we recall that $\bar f_1^{(0)}$ is independent of $s$).

For comparison with (\ref{eq20}), the analogous non-propagating
normal modes in a cigar trap at $T=0$ have a spectrum given by
\cite{zaremba,fl} 
\begin{equation}
\omega_l^2=2l(l+1)\omega_0^2,\ l=0,1,2,...
\label{eq22}
\end{equation}

\section{propagating modes}

In this section, we discuss the more interesting propagating solution
($\bar k\neq0$) of equations (12a) and (12b).
In the classical limit ($B(z_0)=1$),
one immediately finds a phonon mode solution
\begin{equation}
\bar h=\exp(2s/5),\ \ \bar f=0,\ \ \bar\omega^2=\frac{5}{6}\bar k^2.
\label{eq23}
\end{equation}
This is a longitudinal sound wave with the dispersion relation $\omega=c_0k$, 
where the sound velocity $c_0^2=5k_{\rm B}T/3m$ is the same as for a
uniform classical gas.
There is no radial oscillation (i.e. $\bar f=0$) associated with
this phonon mode in the classical limit.
Using (\ref{eq14}), the associated density fluctuation is found to be
\begin{equation}
\delta n({\bf r},t) \propto \exp(-3r_{\perp}^2/5R^2)
\exp(ikz-ic_0kt).
\label{eq24}
\end{equation}
In the classical limit, one can show that the dispersion relation
$\omega=c_0k$
is valid for any cylindrical trap potential $U_0(r_{\perp})$,
i.e. it is not limited to parabolic potentials.
In this more general case, the solution of the hydrodynamic equation 
corresponding to the phonon-like mode in the classical limit is given by
$h=\exp(2U_0/5k_{\rm B}T)$ and $f=0$,
with the associated density fluctuation
$\delta n\propto \exp(-3U_0/5k_{\rm B}T)$.

In order to find how the non-propagating normal mode solutions in (17)
are modified when $k\neq0$, we expand $\bar f$ as follows:
\begin{equation}
\bar f(s)=\sum_n a_n \bar f_n^{(0)}(s).
\label{eq25}
\end{equation}
This follows the approach of Zaremba~\cite{zaremba} for a Bose-condensed
gas at $T=0$ in a cigar-shaped trap.
Substituting (\ref{eq25}) into (12a) and (12b), we obtain the coupled linear
equations for the coefficients $a_n$:
\begin{equation}
\left( \bar\omega^2-\left[\bar\omega_n^{(0)}\right]^2
-\frac{5}{6}\frac{\left[\bar\omega_n^{(0)}\right]^2}
{\bar\omega^2-\frac{5}{6}\bar k^2} \bar k^2 \right) a_n
+\frac{2}{3}\frac{\bar k^2}{\bar\omega^2-\frac{5}{6}\bar k^2}\sum_{n'}
M_{nn'}a_{n'}=0,
\label{eq26}
\end{equation}
where the matrix elements $M_{nn'}$ are defined by
\begin{equation}
M_{nn'}\equiv\int_0^{\infty}ds\ s^2e^{-s}\bar f_n^{(0)}(s)
\bar f_{n'}^{(0)}(s).
\label{eq27}
\end{equation}
Using the identity for Laguerre polynomials
\begin{equation}
\int_0^{\infty}ds\ e^{-s}L_n(s)L_{n'}(s)=\delta_{nn'}(n!)^2,
\label{eq28}
\end{equation}
we find
\begin{equation}
M_{nn'}=2n\delta_{nn'}-\sqrt{nn'}\delta_{n',n\pm1}.
\label{eq29}
\end{equation}
To lowest order in $\bar k^2$, one finds the eigenvalue $\bar\omega^2$ of
(\ref{eq26}) is given by ($M_{nn}=2n$)
\begin{equation}
\bar \omega^2\simeq\frac{10}{3}n+\left(\frac{5}{6}-\frac{M_{nn}}{5n}\right)
\bar k^2=\frac{10}{3}n+\frac{13}{30}\bar k^2.
\label{eq30}
\end{equation}
In ordinary units, the excitation spectrum is given by
\begin{equation}
\omega_n^2(k)=n\frac{10}{3}\omega_0^2+ \frac{13}{15}\frac{k_{\rm B}T}{m}k^2,
\ \ n=1,2,3,\cdots.
\label{eq31}
\end{equation}
We note that the correction term in (\ref{eq31}) is of order $(kR)^2$
relative to the first term, which is assumed to be large.
Thus this spectrum for propagating modes for a classical gas in a cylindrical
harmonic trap is only valid for $kR\ll 1$,
where $R$ is the radial size of the trapped
gas density profile.

\section{Variational solutions}

For a Bose gas above $T_{\rm BEC}$, one cannot easily solve the coupled
equations in (12a) and (12b) for $k\neq0$.
An alternative approach is to recast these equations into a variational
form, following recent work~\cite{zg} in solving the two-fluid 
hydrodynamic equations
for a trapped Bose-condensed gas~\cite{zgn}.
One finds that the functional 
\begin{equation}
E[\bar f,\bar h]\equiv \frac{\displaystyle
{\rm Re}\int_0^{\infty} ds\ g_{3/2}(z_0)
\left[s\bar f^*\hat L[\bar f]+\frac{5}{6}B(z_0)|\bar h|^2\bar k^2
-2s\bar f^*\left(\frac{5}{3}B(z_0)\frac{d\bar h}{ds}-\frac{2}{3}\bar h
\right)\bar k\right] }
{\displaystyle \int_0^{\infty} ds\ g_{3/2}(z_0)(s|\bar f|^2+|\bar h|^2)}.
\label{eq32}
\end{equation}
has the property that conditions $\delta E/\delta\bar f=0,
\delta E/\delta\bar h=0$
yields Eqs.(12a) and (12b).
Thus the normal mode eigenvalues $\omega^2$ are given by the stationary
value of this
functional $E[\bar f,\bar h]$.

For $\bar k=0$ with $\bar \omega\neq0$, we can use (\ref{eq32}) to estimate
the normal mode frequencies using the classical solutions of (\ref{eq17}),
$\bar f=\bar f^{(0)}_n$ and $\bar h=0$,
as trial functions.
The frequency so determined is given by
\begin{eqnarray}
\bar \omega^2&=&\frac{\displaystyle
\int_0^{\infty} ds\ g_{3/2}(z_0)s \bar f^{(0)}_n\hat L[\bar f^{(0)}_n]}
{\displaystyle \int_0^{\infty} ds\ g_{3/2}(z_0)s(\bar f^{(0)}_n)^2} \cr
&=&\frac{10}{3}n+\frac{10}{3}
\frac{\displaystyle \int_0^{\infty} ds\ [g_{3/2}(z_0)-g_{5/2}(z_0)]
\bar f_n^{(0)}\left(s^2\frac{d^2f_n^{(0)}}{ds^2}
+2s\frac{d\bar f_n^{(0)}}{ds}\right)}
{\displaystyle \int_0^{\infty}ds\ g_{3/2}(z_0)s(\bar f_n^{(0)})^2}.
\label{eq33}
\end{eqnarray}
The second term in (\ref{eq33}) gives the quantum correction to the
classical limit.

For $k\neq0$, the most interesting propagating mode is the phonon mode
with $\omega\propto k$.
For trial functions in (\ref{eq32}), we take (see (\ref{eq23})) 
\begin{equation}
\bar f=\bar k A_f,\ \ \bar h=A_h\exp(2s/5),
\label{eq34}
\end{equation}
where the constants $A_f$ and $A_h$ are real and independent of $s$.
To first order in $\bar k$, we find a phonon-like solution 
$\bar \omega=\bar c \bar k$, with the (dimensionless) sound velocity
$\bar c$ given by
\begin{equation}
\bar c^2=\frac{\displaystyle \left(\frac{5}{6}\int_0^{\infty}
ds \ g_{5/2}(z_0)e^{4s/5}-\frac{2}{15}
\left\{\int_0^{\infty} ds[g_{3/2}(z_0)-g_{5/2}(z_0)]se^{2s/5}\right\}^2/
\int_0^{\infty} ds\ g_{3/2}(z_0)s \right)}
{\displaystyle \int_0^{\infty} ds\ g_{3/2}(z_0)e^{4s/5}}.
\label{eq35}
\end{equation}
The amplitudes in (\ref{eq34}) which are associated with this phonon mode
have the ratio
\begin{equation}
\frac{A_f}{A_h}=-\frac{\displaystyle \int_0^{\infty} ds
[g_{3/2}(z_0)-g_{5/2}(z_0)]se^{2s/5}}
{\displaystyle 5\int_0^{\infty} ds\ g_{3/2}(z_0)s}.
\label{eq36}
\end{equation}
One can see that in a degenerate Bose gas, where
$g_{3/2}(z_0)\neq g_{5/2}(z_0)$, 
the {\it radial} oscillations ($A_f$) are coupled to the {\it axial}
oscillations ($A_h$).

The normal mode frequencies given by the variational expressions
in (\ref{eq33}) and (\ref{eq35}) can be numerically calculated.
All the integrals can be evaluated analytically using the useful
identity,
\begin{equation}
\int_0^{\infty}ds\ g_n(z_0)s^m=m!g_{n+m+1}(\tilde z_0),
\label{37}
\end{equation}
where $\tilde z_0\equiv e^{\beta\mu_0}$ and $g_n$ is the Bose-Einstein
function, as defined below (2).
It is useful to plot the temperature dependence relative to the
$T_{\rm BEC}$
for an ideal gas in a cigar-shaped trap described by (3).
For a trap of length $L$ with $N$ atoms, we have (using (2))
\begin{eqnarray}
N=\int d{\bf r}\ n_0({\bf r})&=&\frac{2\pi L}{\Lambda^3}\int_0^{\infty}
dr_{\perp}\ r_{\perp}g_{3/2}(z_0) \cr
&& \cr
&=&\frac{\pi R^2L}{\Lambda^3}\int_0^{\infty}ds\ g_{3/2}(\tilde z_0e^{-s}) \cr
&& \cr
&=&L\left(\frac{m\omega_0}{2\pi\hbar}\right)^{1/2}
\left(\frac{k_{\rm B}T}{\hbar\omega_0}\right)^{5/2}g_{5/2}(\tilde z_0).
\label{eq38}
\end{eqnarray}
When $T=T_{\rm BEC}$, we have $\mu_0=0$ and hence $\tilde z_0=1$.
Thus (\ref{eq38}) gives the Bose-Einstein transition temperature
\begin{equation}
k_{\rm B}T_{\rm BEC}=\hbar\omega_0\left[\frac{N}{L}
\left(\frac{2\pi\hbar^2}{m\omega_0}\right)^{1/2}\frac{1}{\zeta(5/2)}
\right]^{2/5}
\label{eq39}
\end{equation}
for a cigar trap in the usual semiclassical approximation.

In Fig.~1, we show how the sound velocity given by (\ref{eq35})
varies with temperature down to $T_{\rm BEC}$,
relative to the classical value $c_0=\sqrt{5k_{\rm B}T/3m}$.
In Fig.~2, we show the temperature-dependent results for the frequencies
of the non-propagating modes based on (\ref{eq33}).
We recall that since $f_1^{(0)}$ is constant, the correction term in
(\ref{eq33}) vanishes for the $n=1$ mode.
The variational calculations shown in Figs.~1 and 2 indicate that there is
little change in the normal mode frequencies given by the
classical limit for temperature down to $T=T_{\rm BEC}$.
As noted at the end of Ref.\cite{gws}, these results are to be expected
since the only place where the Bose nature of the gas enters is in the first
term of Eq.~(\ref{eq1}).
This involves the ratio $B(z_0)=g_{5/2}(z_0)/g_{3/2}(z_0)$, which is
remarkably close to the classical value of unity, even at the center
of the trap.

\section{propagation of sound pulses}
In this section, following the approach of Zaremba~\cite{zaremba},
we discuss the propagation of sound pulses 
induced by a small external perturbation $\delta U({\bf r},t)$.
We assume that $\delta U({\bf r},t)$ has no radial dependence and is
switched on at $t=0$, i.e. the external perturbation is of the form
\begin{equation}
\delta U({\bf r},t)=\delta U(z)\theta(t).
\label{eq40}
\end{equation}
The equation of motion (\ref{eq1}) with the external perturbation
$\delta U$ can
be solved~\cite{zaremba} by introducing a Fourier representation of the
velocity fluctuations (compare with (\ref{eq5})) and the external perturbation
\begin{eqnarray}
{\bf v}({\bf r},t)&=&\int_{-\infty}^{\infty}\frac{dk}{2\pi}e^{ikz}
(xf(k, r_{\perp},t),yf(k, r_{\perp},t),h(k,r_{\perp},t)),\cr
\delta U(z)&=&\int_{-\infty}^{\infty}\frac{dk}{2\pi}e^{ikz}\delta U(k).
\label{eq41}
\end{eqnarray}
Taking the Fourier transform of (\ref{eq41}) and using the radial variable $s$
defined in (\ref{eq9}), we obtain a coupled set of equations
(compare with (12a) and (12b))
\begin{mathletters}
\begin{eqnarray}
&&\frac{\partial^2 \bar f}{\partial t^2}+\omega_0^2\left[\hat L[\bar f]
-\bar k \left(\frac{5}{3}B(z_0)\frac{\partial \bar h}{\partial s}
-\frac{2}{3}\bar h \right)\right]=0, \\
&&\frac{\partial^2 \bar h}{\partial t^2}+\omega_0^2\left\{
\frac{5}{6}B(z_0)\bar k^2
\bar h^2- \bar k\left[\frac{5}{3}B(z_0)s\frac{\partial \bar f}
{\partial s}+\left(\frac{5}{3}B(z_0)-s\right)\bar f\right]\right\}
=-i\frac{k}{m}\delta U(k)\delta(t).
\label{eq42}
\end{eqnarray}
\end{mathletters}

\noindent
Here we have used notation analogous to that in (\ref{eq11}), but now 
$\bar f$ and $\bar h$ also depend on $t$ and are for particular $k$
component.

In order to solve these coupled equations, we expand $\bar f$ and $\bar h$ as
follows,
\begin{equation}
\bar f(k,s,t)=\sum_m b_m(k,t)\bar f_m(k,s),
\ \ \bar h(k,s,t)=\sum_m b_m(k,t)\bar h_m(k,s),
\label{eq43}
\end{equation}
where the basis functions $\bar f_m$ and $\bar h_m$ are the
normal mode solutions of (12).
These have eigenvalues
$\bar\omega_m^2(k)$ and form an orthonormal set (see (8b)),
\begin{equation}
\int_0^{\infty} ds\ g_{3/2}(z_0)[s\bar f_{m}^*\bar f_n+\bar h_{m}^* \bar h_n]
=\delta_{mn}.
\label{eq44}
\end{equation}
Substituting the expression (\ref{eq43}) into (42) and using
(\ref{eq44}), we obtain
\begin{equation}
\frac{\partial^2b_n}{\partial t^2}+\omega_n^2(k)b_n=F_n(k)\delta(t),
\label{eq45}
\end{equation}
where $\omega_n(k)\equiv\bar\omega_n(k)\omega_0$ and
\begin{equation}
F_n(k)\equiv -ik\frac{\delta U(k)}{m}\int_0^{\infty} ds\ g_{3/2}(z_0){\bar h}
^*_n(k,s).
\label{eq46}
\end{equation}
The solution of (\ref{eq45}) with the boundary condition $b_n(t<0)=0$ is
given by
\begin{equation}
b_n(k,t)=\theta(t)\frac{F_n(k)}{\omega_n(k)}\sin[\omega_n(k)t],
\label{eq47}
\end{equation}
where the quantum number $n$ refers to the radial degree of freedom.

In order to analyse the time evolution of the associated density fluctuations,
it is convenient to work with the radially-averaged density
\begin{equation}
\overline{\delta n}(z,t)\equiv\int d{\bf r}_{\perp}\delta n({\bf r},t).
\end{equation}
Using (\ref{eq14}), one finds
\begin{eqnarray}
\frac{\partial}{\partial t}\overline{\delta n}&=&-\int d{\bf r}_{\perp}
\bbox{\nabla} \cdot(n_0{\bf v})=-\int d{\bf r}_{\perp}n_0\frac{\partial v_z}
{\partial z} \cr
&& \cr
&=&-i\frac{2\pi}{\Lambda^3}\int_0^{\infty}r_{\perp}dr_{\perp}\ g_{3/2}(z_0)
\int_{-\infty}^{\infty}\frac{dk}{2\pi}ke^{ikz}h(k,r_{\perp},t) \cr
&& \cr
&=&-\frac{\pi R^2}{\Lambda^3}\sum_n\int_{-\infty}^{\infty}
\frac{dk}{2\pi}e^{ikz}
\frac{k^2\delta U(k)}{m\omega_n(k)}\theta(t)\sin \omega_n(k)t
\left|\int_0^{\infty} ds g_{3/2}(z_0)\bar h_n(k,s)\right|^2.
\label{eq49}
\end{eqnarray}
Integrating (\ref{eq49}) over $t$, one finds
\begin{equation}
\overline{\delta n}(z,t)
=-\frac{\pi R^2}{\Lambda^3}\sum_n\int_{-\infty}^{\infty}
\frac{dk}{2\pi}e^{ikz}
\frac{k^2\delta U(k)}{m\omega^2_n(k)}\theta(t)[1-\cos \omega_n(k)t]
\left|\int_0^{\infty} ds\  g_{3/2}(z_0)\bar h_n(k,s)\right|^2.
\label{eq50}
\end{equation}
One can see that a low frequency phonon mode makes a large contribution
to (\ref{eq50}).
To illustrate the contribution to (\ref{eq50}) which is associated with the
phonon density fluctuations $\omega_n=ck$ in the classical limit,
we use $\bar h_n(k,s)=A_h\exp(2s/5)$ (see (\ref{eq23})) where the
normalization condition (44) gives $A_h^2=1/5\tilde z_0$.
The contribution to (\ref{eq50}) of this classical sound wave is given
by (we use $\frac{N}{L}=\frac{\pi R^2}{\Lambda^3}\tilde z_0$ appropriate
to the classical limit)
\begin{equation}
\overline {\delta n}(z,t)=-\frac{5}{9}\frac{N}{L}\frac{\theta(t)}{mc^2}
\left\{\delta U(z)-\frac{1}{2}[\delta U(z-ct)+\delta U(z+ct)]\right\},
\label{eq51}
\end{equation}
which has the form of a propagating pulse moving with a speed $\pm c$.

\section{concluding remarks}
In this paper, we have given a detailed analysis of the hydrodynamic normal modes of a Bose gas in a cigar-shaped trap above $T_{\rm BEC}$.
We have discussed the non-propagating and propagating modes, both in the
classical limit as well as in the degenerate Bose limit just above
$T_{\rm BEC}$.
Our results complement the analogous
studies~\cite{zaremba,st2,pethick,machida} 
of such modes in the quantum hydrodynamic limit at $T=0$.
In contrast with the $T=0$ analysis, which works with a single equation
for the density fluctuations, we have to work with coupled equations for
the velocity fluctuations.
For simplicity, we have considered the limit of a uniform gas along the
axial direction.
We note that Stringari~\cite{st2} has discussed the condensate modes at
$T=0$ in the limit of a very weak trap in
the axial direction ($\omega_z\ll\omega_0$).

As in Ref.~\cite{gws}, we have ignored the Hartree-Fock mean field
contribution to the hydrodynamic equations.
Such terms are given in Eq.~(6) of Ref.~\cite{zgn}.
In place of (\ref{eq1}), we obtain the linearized velocity equation
\begin{eqnarray}
m\frac{\partial^2{\bf v}}{\partial t^2}&=&
\frac{5P_0({\bf r})}{3n_0({\bf r})}\bbox{\nabla}(\bbox{\nabla}\cdot
{\bf v})
-\bbox{\nabla}[{\bf v}\cdot \bbox{\nabla}U({\bf r})]
-\frac{2}{3}(\nabla\cdot{\bf v})\bbox{\nabla}U({\bf r}) \cr
&&+2g\bbox{\nabla}(\bbox{\nabla}\cdot n_0{\bf v})
-\frac{\partial}{\partial t}\bbox{\nabla}\delta U({\bf r},t).
\label{new52}
\end{eqnarray}
This now involves the effective trap potential
\begin{equation}
U({\bf r})=U_0({\bf r})+2gn_0({\bf r}),
\label{new53}
\end{equation}
which also appears in the equilibrium fugacity
$z_0=e^{\beta(\mu_0-U)}$ in the expressions for $n_0({\bf r})$ and 
$P_0({\bf r})$.
The usual $s$-wave scattering interaction is $g=4\pi a\hbar^2/m$.
The analysis given in this paper can be generalized \cite{nikuni}
to include the effects of this HF mean field but it is much more complicated.
We simply quote some final results for the classical limit.
The $n=1$ non-propagating mode has a frequency given by
\begin{equation}
\omega_1^2=\frac{10}{3}\omega_0^2
\left(1-\frac{gn_0({\bf r}=0)}{2k_{\rm B}T}\right),
\label{neq54}
\end{equation}
where $n_0({\bf r}=0)$ is the density at the center of the cylindrical trap.
The sound velocity corresponding to (23) is given by
\begin{equation}
c^2=\frac{5k_{\rm B}T}{3m}+\frac{gn_0({\bf r}=0)}{3m}.
\label{new55}
\end{equation}

The two-fluid hydrodynamic equations for a trapped Bose-condensed gas
($T<T_{\rm BEC}$) have been recently discussed by Zaremba and the
authors~\cite{zgn}.
These equations have been used to study first and second sound modes in a
dilute uniform Bose gas~\cite{gz} at finite temperatures.
It is found that first sound corresponds mainly to an oscillation of the 
non-condensate, with a velocity given by
\begin{equation}
u_1^2=\frac{5}{3}\frac{k_{\rm B}T}{m}\frac{g_{5/2}(z_0)}{g_{3/2}(z_0)}
+\frac{2g\tilde n_0}{m}.
\end{equation}
In contrast, the second sound mode mainly corresponds to an oscillation
of the condensate, with a velocity given by
\begin{equation}
u_2^2=\frac{gn_{c0}}{m}.
\label{eq53}
\end{equation}
Here $n_{c0}(\tilde n_0)$ is the equilibrium condensate (non-condensate)
density.
As discussed in Ref.~\cite{gz}, to a good approximation, one can use
\begin{equation}
\tilde n_0=\frac{1}{\Lambda^3}g_{3/2}(z_0),
\end{equation}
where the equilibrium fugacity is $z_0=e^{-\beta gn_{c0}}$.

In principle, we could use the equations in Ref.~\cite{zgn} to extend
the analysis of the
present paper and discuss the propagating first and second sound modes in
a cigar-shaped trap.
Here we limit ourselves to some qualitative remarks.
One expects to find an expression similar to (\ref{eq51}) for the propagation
of a pulse, and there should be distinct first and second sound pulses
moving with velocities quite close to $u_1$ and $u_2$ as given above.
However, as the expression in (\ref{eq51}) shows, the relative amplitude of
these two modes is proportional to $1/u_i^2$.
We conclude that if pulse experiments such as in Ref.~\cite{andrews}
were done in the hydrodynamic region, most of the weight would be
in the second sound pulse if $u_2^2\ll u_1^2$.
This mode, given by (\ref{eq53}), is the natural hydrodynamic analogue of the
Bogoliubov mode exhibited in the quantum hydrodynamic region at
$T=0$~\cite{andrews,zaremba}.
At temperatures close to $T_{\rm BEC}$, 
the first sound pulse has a much faster speed and thus 
its intensity will be very weak.
The observation of distinct first and second sound
pulses in cigar-shaped traps would be very dramatic evidence for superfluid
behavior in dilute Bose gases.
The experiment would best be done at intermediate or lower temperatures,
where $u_1$ and $u_2$ are more comparable in magnitude.
\acknowledgments

T.N. was supported by a fellowship from the Japan Society for the Promotion
of Science (JSPS) and  A.G. by a research grant from NSERC of Canada.
A.G. would also like to thank the Institute for Theoretical Physics
at Santa Barbara for support during the final stages of work on this paper.

\bigskip
\bigskip

\centerline{\bf FIGURE CAPTIONS}
\begin{itemize}
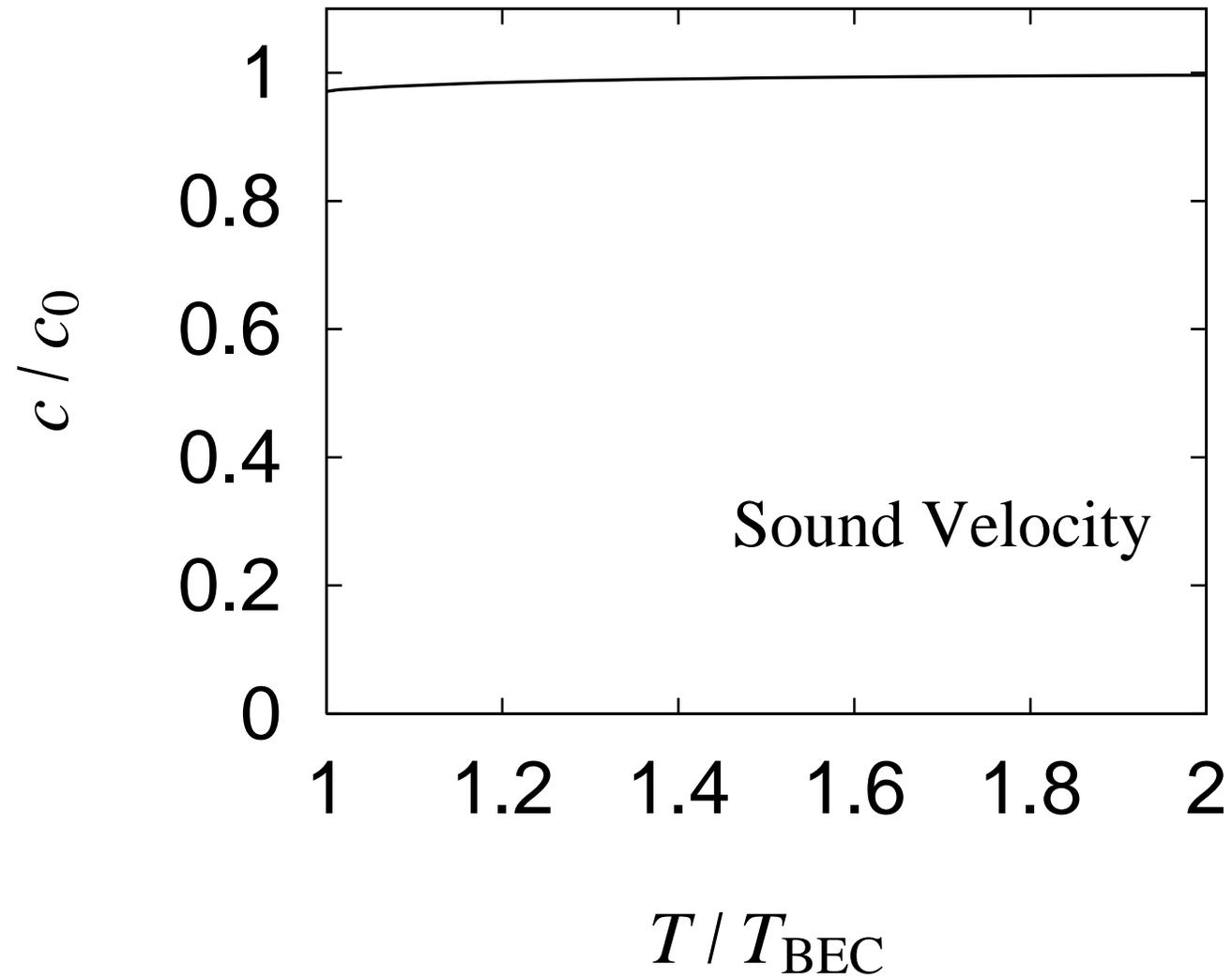
\item[Fig.1:] 
The sound velocity $c$ as a function of temperature relative
to $T_{\rm BEC}$.
The values are normalized to the classical gas result
$c_0=\sqrt{5k_{\rm B}T/m}$.

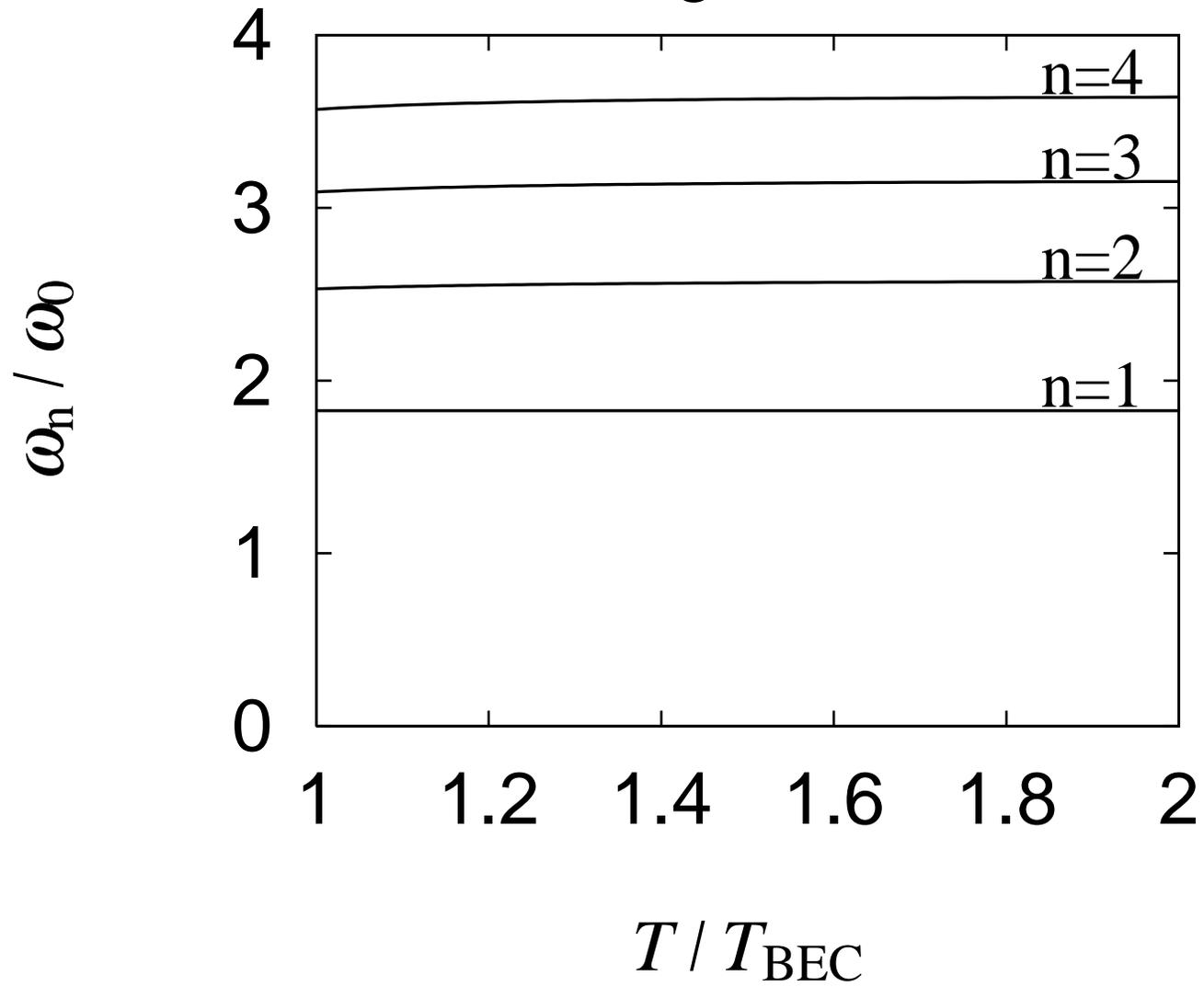
\item[Fig.2:] 
The normal mode frequencies $\omega_n$ for $k=0$ as a function of
temperature, as given by (\ref{eq33}).
The frequencies are normalized to the radial trap frequency
 $\omega_0$.

\end{itemize}

\begin{references}

\bibitem{andrews}M.R. Andrews, D.M. Kurn, H.-J. Miesner, D.S. Durfee,
C.G. Townsend, S. Inouye and W. Ketterle,
Phys. Rev. Lett. {\bf 79}, 553 (1997); {\it ibid}. {\bf 80}, 2967 (1998).

\bibitem{zaremba}E. Zaremba, Phys. Rev. {\bf A57}, 518 (1998).

\bibitem{st1}S. Stringari, Phys. Rev. Lett. {\bf 77}, 2360 (1996).

\bibitem{st2}S. Stringari, cond-mat/9801067.

\bibitem{pethick}G.M. Kavoulakis and C.J. Pethick, cond-mat/971124.

\bibitem{machida}T. Isoshima and K. Machida, cond-mat/9712122.

\bibitem{gws} A. Griffin, W.C. Wu and S. Stringari,
Phys. Rev. Lett. {\bf 78}, 1838 (1997).

\bibitem{kavou} See also G.M. Kavoulakis, C.J. Pethick and H. Smith,
Phys. Rev. A {\bf 57}, 2938 (1998).

\bibitem{fl} M. Fliesser, A. Csord\'as, P. Sz\'epfalusy
and R. Graham, Phys. Rev. A {\bf 56}, R2533 (1997).

\bibitem{zg} E. Zaremba and A. Griffin, to be published.

\bibitem{zgn} E. Zaremba, A. Griffin and T. Nikuni, Phys. Rev. {\bf A57},
4695 (1998).

\bibitem{nikuni} T. Nikuni, unpublished.

\bibitem{gz} A. Griffin and E. Zaremba, Phys. Rev. {\bf A56}, 4839
(1997).

\end{references}
\end{document}